\documentclass[aps,twocolumn,pre]{revtex4}
\usepackage{graphpap}
\usepackage[dvips]{graphicx}
\usepackage[dvips]{graphics}
\usepackage{color}

\begin{document}

\title{Comment on "Dynamic range of nanotube- and nanowire-based
  electromechanical systems" [APL 86, 223105 (2005)]} 
 \author{Christoph Stampfer}
 \affiliation{Micro-~and Nanosystems, Department of Mechanical and Process
   Engineering, ETH Z\"urich, Tannenstrasse 3, CH-8092 Z\"urich, Switzerland.}
 \author{Stefan Rotter and Joachim Burgd\"orfer}
 \affiliation{Institute for Theoretical Physics, Vienna University of
   Technology, A-1040 Vienna, Austria.} 
 \date{ \today}
 \begin{abstract}
 \end{abstract}
 \maketitle
The field of nano-electromechanical systems (NEMS) has attracted great
interest in the last decade and a variety of different applications of NEMS
(such
as ultra-sensitive force and mass detectors) have been proposed. Very recently,
an interesting work on the dynamic range of nanotube and
nanowire-based NEMS has been presented \cite{pos05}.\\The aim of this comment
is to emphasize quantum effects in the
dynamic range (DR) of nanoscaled resonators. 
We demonstrate that the zero-point
energy contribution starts to play a significat role near the
temperature and size regimes considered in Ref.~\cite{pos05}.\\A
crucial quantity in the derivation of the DR (see Eq.~(8) in
Ref.~\cite{pos05}) is the thermo-mechanical noise, expressed by the spectral
density of the displacement noise (on resonance $\omega_0$) given by (see 
Eq.~(7) in Ref.~\cite{pos05}),
\begin{equation}\label{eq1}
S_x=4 k_B T \frac{Q}{m \omega_0^3}\,,
\end{equation}
which is based on the classical equipartition law ($k_BT \gg \hbar\omega_0$)
and on the classical fluctuation-dissipation theorem (CFDT). 
\begin{table}[b]
\begin{center} {\footnotesize
\begin{tabular}{|c||c|c|c|c|}
\hline
${\displaystyle \hbar \omega_0/(k_B T) }$ & {$l=200nm$} &
{$l=100nm$} & {$l=50nm$} & $l=25nm$  \\
\hline\hline
SWNT &    0.008 (0.024)  &  0.03 (0.098)  &   0.13 (0.39) & 0.55 (1.57)\\\hline
MWNT  &    0.06  & 0.25  &   0.99  &  3.96    \\\hline
Pt nanowire &   0.03  &   0.13  &   0.53 &  2.11   \\\hline
SiC resonator &   0.57  &   2.30  &   9.21 &  36.8   \\
\hline
\end{tabular} }
\end{center}
\caption{Ratio $\hbar \omega_0 /(k_B T) $ for several doubly clamped resonators
  at $4 K$. 
  The results for the SWNT are calculated with the model of a solid cylinder
  (as in Ref.~\cite{pos05}) and, alternatively, with the model of a hollow
  cylinder (see results in brackets):  
  diameter $d=1.4~nm$, section area $A=\pi d t$, moment of inertia 
  $I=d^3 t/8$, and the layer spacing for graphite $t = 0.34 nm$.}
\label{turns}
\end{table}
However, since the systems discussed in Ref.~\cite{pos05} 
will be driven at low temperatures (see the ratios of $\hbar \omega_0 /
(k_B T) $ in Table I) quantum fluctuations
should be taken into account in the thermo-mechanical noise. 
\begin{figure}[t]\centering
\includegraphics[draft=false,keepaspectratio=true,clip,%
                   width=0.85\linewidth]%
                   {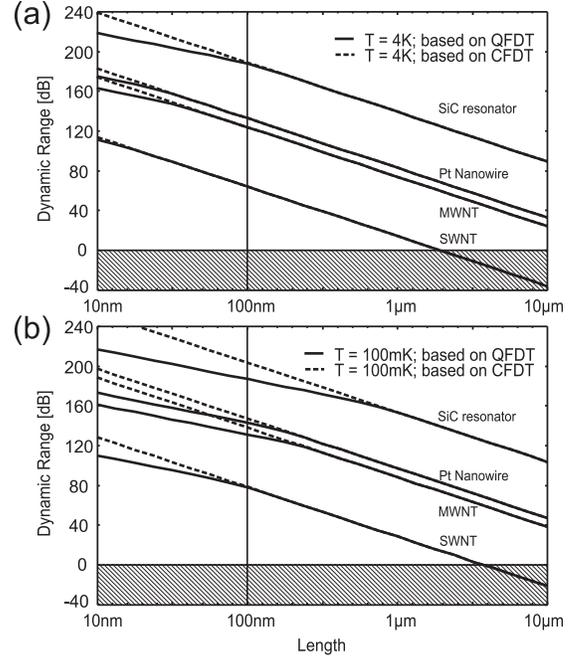}                   
\caption{Dynamic range (DR) as a function of resonator length at (a) 
$4 K$ and (b) $100mK$ for several doubly clamped resonators based
on the classical and quantum fluctuation dissipation theorem (C/QFDT). 
Note the significant discrepancy between the DR based on CFDT vs.~QFDT and
the expected temperature independence of the QFDT-results 
for short resonator length.}
\label{trdansport}
\end{figure} 
The most direct way to include the quantum noise is to substitute
the thermal energy $k_B T$ by the thermally
averaged quantum energy including the zero point energy contribution
(following Callen and Welton \cite{cal01}). 
Employing the quantum
fluctuation-dissipation theorem (QFDT), the expression for the spectral
density of the displacement noise (again on resonance $\omega_0$) is thus
given by (see Ref.~\cite{hut93} and references therein)
\begin{equation}\label{eq2}
S_x =  2 \hbar \omega_0 \coth \left( \frac{\hbar \omega_0}{2 k_B T} \right)
\frac{Q}{m \omega_0^3}\,.
\end{equation}
In the high temperature limit ($k_B T \gg \hbar \omega_0$) 
we recover Eq.~(\ref{eq1}) corresponding to the results of
Nyquist and Johnson. Substituting Eq.~(\ref{eq2}) 
into the expression for the DR (Eq.~(8) in Ref.~\cite{pos05}) 
leads to the modified DR based on
QFDT. A replot of Fig.~3 in Ref.~\cite{pos05} is shown in 
Fig.~\ref{trdansport}, where the DR based on the CFDT and the QFDT
are plotted as a function of the resonator length.
Note the significant discrepancy between the classical and the quantum 
DR in temperature and length regimes accessible
by the experiment. 
The numerical data shown in Fig.~\ref{trdansport} and in Table I have 
been computed using Eq.~(9) and the parameters given in Table I of
Ref.~\cite{pos05}. To reproduce Fig.~3 of Ref.~\cite{pos05} we, however, 
choose the diameter of the MWNT as $d=10nm$ and the Q-factor of the
SiC-beam as $Q=4500$ (as in Ref.~[18] of Ref.~\cite{pos05}).\\In conclusion,
we have shown that the quantum fluctuation-dissipation relation
poses limits on the dynamics of nanoscaled
mechanical resonators and cannot be neglected in a description of 
NEMS at low temperatures.\\C.~S.~acknowledges support 
by the Micro- and Nanosystems Chair and the grant TH-18/03-1.

\end{document}